# Se content $x$ dependence of electron correlation strength in Fe$_{1+y}$Te$_{1-x}$Se$_x$

L. C. C. Ambolode II[1], K. Okazaki[1], M. Horio[1], H. Suzuki[1], L. Liu[1], S. Ideta[1], T. Yoshida[1], T. Mikami[1], T. Kakeshita[1], S. Uchida[1], K. Ono[2], H. Kumigashira[2], M. Hashimoto[3], D.-H. Lu[3], Z.-X. Shen[4,5], and A. Fujimori[1]

[1]*Department of Physics, University of Tokyo, Tokyo 113-0033, Japan*
[2]*Photon Factory, Institute of Materials Structure Science, KEK, Tsukuba, Ibaraki 305-0801, Japan*
[3]*Stanford Synchrotron Radiation Lightsource, SLAC National Accelerator Laboratory, Menlo Park, CA 94025, USA*
[4]*Stanford Institute of Materials and Energy Sciences, SLAC National Accelerator Laboratory, Menlo Park, CA 94025, USA*
[5]*Departments of Physics and Applied Physics, and Geballe Laboratory for Advanced Materials, Stanford University, Stanford, CA 94305, USA*

The iron chalcogenide Fe$_{1+y}$Te$_{1-x}$Se$_x$ on the Te-rich side is known to exhibit the strongest electron correlations among the Fe-based superconductors, and is non-superconducting for $x < 0.1$. In order to understand the origin of such behaviors, we have performed ARPES studies of Fe$_{1+y}$Te$_{1-x}$Se$_x$ ($x = 0$, 0.1, 0.2, and 0.4). The obtained mass renormalization factors for different energy bands are qualitatively consistent with DFT + DMFT calculations. Our results provide evidence for strong orbital dependence of mass renormalization, and systematic data which help us to resolve inconsistencies with other experimental data. The unusually strong orbital dependence of mass renormalization in Te-rich Fe$_{1+y}$Te$_{1-x}$Se$_x$ arises from the dominant contribution to the Fermi surface of the $d_{xy}$ band, which is the most strongly correlated and may contribute to the suppression of superconductivity.

**Introduction**

Although all the iron pnictide and iron chalcogenide superconductors share the same Fe-pnictogen/chalcogen layers [1], significant variations have been observed in their physical properties such as ordered magnetic moments, effective band masses, superconducting gaps, and transition temperatures [2]. In the cuprates, strong electron correlations play a vital role in their unusual physical properties while it is still unclear as to what extent electron correlations affect the physical properties including the superconductivity of the iron-based superconductors. From the theoretical side, combined density functional theory and dynamical mean-field theory [3] (DFT + DMFT) studies have addressed this issue [2, 4]. FeTe$_{1-x}$Se$_x$, so-called 11 system, has the simplest crystal structure among the iron-based superconductors, consisting only of FeSe/FeTe layers without intervening layers found in the other families [5-7]. Superconductivity occurs between $x = 0.1$ and $x = 1$, as shown in Fig. 1 [8, 9]. Thus, FeTe$_{1-x}$Se$_x$ is an ideal system to gain deeper insight into the origin of the superconductivity and how electron correlations influence it. According to the DFT + DMFT calculation [2], the end member FeTe is predicted to exhibit the strongest electron correlations and strongest orbital dependence among the iron-based superconductors. FeSe, on the other hand, shows only moderate electron correlations and orbital dependence, comparable to those in the other iron-based superconductors. It is also interesting to note that FeSe is a superconductor [10-12] while FeTe is an antiferromagnetic metal [13, 14]. It seems that intermediate correlation strength and large orbital degeneracy are required for superconductivity, and that too strong electron correlations and orbital differentiation may deteriorate superconductivity as seen for FeTe [2, 15]. The obvious difference between FeTe and FeSe is the chalcogen height, the distance between the chalcogen (Te/Se) plane and the Fe plane. This is because the Te atom has a larger radius than the Se atom. Alloying FeTe with FeSe gives us deeper insight into how the strength of electron correlations and its orbital dependence evolve with chalcogen height.

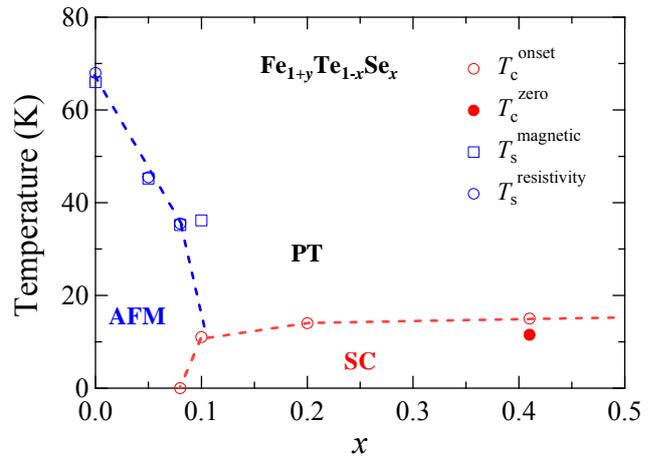

FIG. 1. Phase diagram of Fe$_{1+y}$Te$_{1-x}$Se$_x$ based on in-plane resistivity and magnetic susceptibility measurements of samples used in the present study and of a few additional Te-rich samples [9].

So far, several angle-resolved photoemission spectroscopy (ARPES) studies have been reported on the FeTe$_{1-x}$Se$_x$ compounds. However, there has not been general consensus on the systematic evolution of the electronic structure and orbital-dependent electron correlation strength, and sometimes contradicting results have been reported. From the FeTe end to $x$ = 0.3, two band dispersions were observed around the zone center $\Gamma$ point, where the bands are uniformly renormalized with a factor of about 2 to 3 [15-17]. For $x$ = 0.34, three bands were observed with a uniform mass renormalization of about 3 [18]. From $x$ = 0.34 to the FeSe end, all the three bands were observed, but the mass renormalization exhibited strong orbital dependence [19-21]. To understand the origin of those apparently inconsistent behaviors, we have performed a systematic composition dependent ARPES study of Fe$_{1+y}$Te$_{1-x}$Se$_x$ with $x$ = 0, 0.1, 0.2, and 0.4.

**Methods**

Single crystals of Fe$_{1+y}$Te$_{1-x}$Se$_x$ with nominal Se concentrations of $x$ = 0, 0.1, 0.2 and 0.4 were synthesized using the Bridgman method. The obtained crystals were characterized by in-plane resistivity and magnetic susceptibility measurements and plotted as a phase diagram in Fig. 1. Table 1 shows the result of composition analysis by energy-dispersive x-ray spectroscopy (EDX) and the critical temperature $T_c$ for each sample. More details of the sample characterization are described in Ref. [9]. ARPES experiments were performed at beamline 5-4 of Stanford Synchrotron Radiation Lightsource (SSRL) using a VG-Scienta R4000 energy analyzer. Photoemission data were taken using photons with the energy of $h\nu$ = 22 eV at $T$ = 80, 40, 20, and 9 K for $x$ = 0, 0.1, 0.2, and 0.4 samples, respectively, to focus only on the paramagnetic normal states of each sample (see Fig. 1 for the phase diagram), although the strength of electron correlations may vary with temperature particularly due to finite Hund's coupling [22]. Any signature of a superconducting gap was not recognized for the spectra of $x$ = 0.4 samples taken at 9 K and hence all the present data can be regarded as representing the electronic structure of the normal state. The samples were cleaved *in situ* and measured under a pressure better than 3 × 10$^{-11}$ Torr.

**Table 1.** Chemical compositions determined by energy-dispersive x-ray spectroscopy (EDX) for each nominal Se concentration $x$. The $T_c$ of each sample is also shown.

| $x$ | Fe | Te | Se | $T_c$ |
|---|---|---|---|---|
| 0.4 | 1 | 0.59 | 0.41 | 15.0 |
| 0.2 | 1.06 | 0.8 | 0.2 | 13.9 |
| 0.1 | 1.09 | 0.9 | 0.1 | 10.9 |
| 0 | 1.08 | 1 | 0 | - |

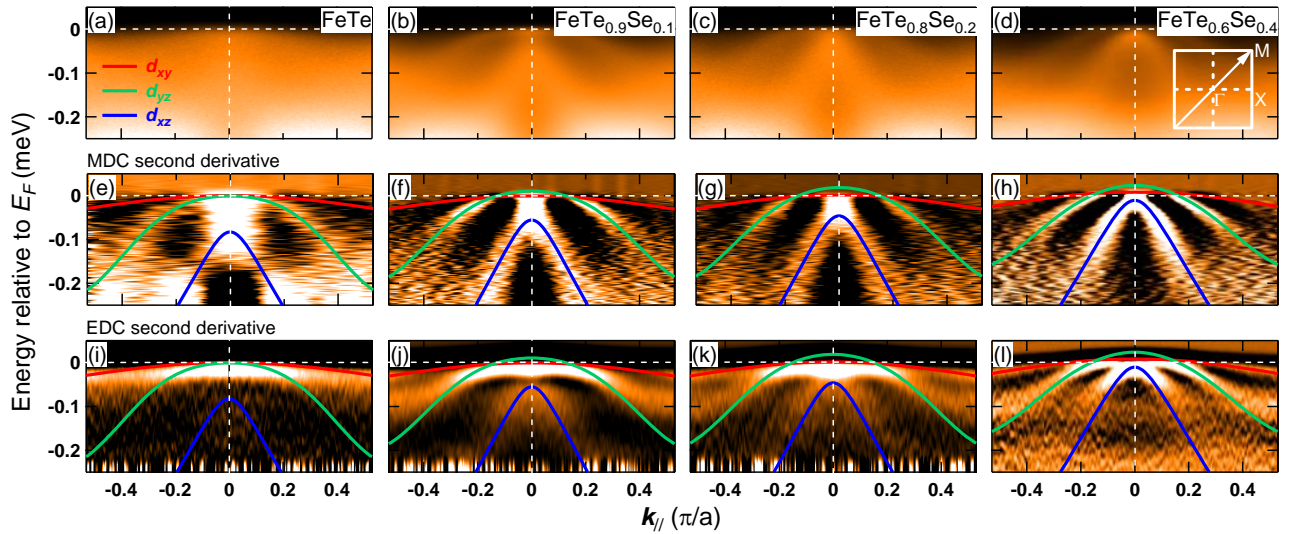

FIG. 2. ARPES spectra of Fe$_{1+y}$Te$_{1-x}$Se$_x$ ($x$ = 0, 0.1, 0.2, and 0.4). (a)-(d) Intensity plots near $E_F$ along the $\Gamma$-M direction. (e)-(h) Second-derivative plots with respect to momentum. (i)-(l) Second-derivative plots with respect to energy. The overlaid lines are calculated bands of $d_{xy}$, $d_{yz}$, and $d_{zx}$ characters scaled and shifted (as indicated in Table 2) so that the best fit to the experimental band dispersions is obtained. Inset of panel (d) is the two-dimensional first Brillouin zone in the $k_z$ = 0 plane.

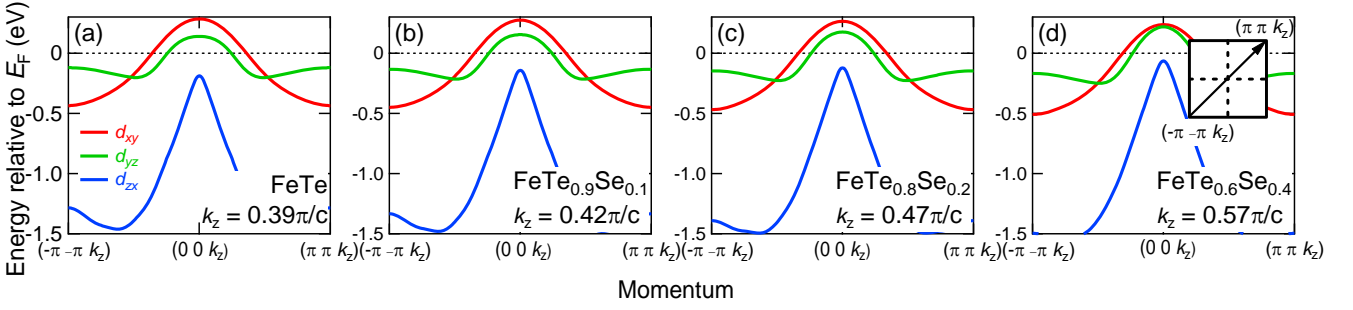

FIG. 3. DFT band structures of "FeTe$_{1-x}$Se$_x$" for various $k_z$ values (0.39, 0.42, 0.47 and 0.57 $\pi/c$) corresponding to the $k_z$ values probed by the ARPES measurements with $h\nu = 22$ eV for the different compositions of Fe$_{1+y}$Te$_{1-x}$Se$_x$ ($x$ = 0, 0.1, 0.2 and 0.4). The chalcogen height was taken from Ref. [6] for FeTe and FeTe$_{0.6}$Se$_{0.4}$, and linearly interpolated between them for FeTe$_{0.9}$Se$_{0.1}$ and FeTe$_{0.8}$Se$_{0.2}$.

In order to compare the ARPES spectra with band theory, we have performed DFT band-structure calculations for FeTe and "FeTe$_{1-x}$Se$_x$" using a WIEN2k package [23]. Calculation was performed on FeTe where the structure data of FeTe$_{1-x}$Se$_x$ were used. The lattice parameters for each composition were taken from Ref. [13] and the chalcogen height was taken from Refs. [6, 24] for FeTe and FeTe$_{0.6}$Se$_{0.4}$, and was linearly interpolated between them for FeTe$_{0.9}$Se$_{0.1}$ and FeTe$_{0.8}$Se$_{0.2}$.

**Results and discussion**

Figures 2(a), (e), and (i) show the ARPES spectra of Fe$_{1.08}$Te ($x$ = 0) measured along the Γ-M line of the two-dimensional Brillouin zone. Even though the spectra are broad in the raw data due to strong quasiparticle scattering originating from spin fluctuations [17], one can clearly observe two band dispersions from the second-derivative spectra of momentum distribution curves (MDCs) [Fig. 2(e)]. The third, weak, less dispersive band near the Fermi level ($E_F$) is also discernable in the second-derivative spectra of energy distribution curves (EDCs) [Fig. 2(i)]. No clear $E_F$ crossing is observed.

Following the assignment of the orbital character of the energy bands by Chen *et al.* based on the polarization-dependent ARPES measurements of FeTe$_{0.66}$Se$_{34}$ [18] and the present band structure calculations, we assign the orbital character of the inner, middle, and outer bands to $d_{zx}$, $d_{yz}$ and $d_{xy}$, respectively. For the sake of comparison with the DFT band-structure calculations (Fig. 3), we have fitted the calculated band structures to the experimental ones as shown by solid curves in Figs. 2(e)-(l). Here, for each energy band, we have rescaled the band dispersion uniformly with the $E_F$ fixed followed by an energy shift to reproduce the experimental band dispersions for each composition. The resulting renormalization factors and the amount of the energy shifts are summarized in Table 2. Contrary to the previous photoemission results [15, 17, 18], the mass renormalization exhibits a systematic orbital dependence. In particular, our results consistently reveal that the $d_{xy}$ band is most strongly renormalized for every composition in the range of 0 < x < 0.4.

**Table 2.** Mass renormalization and energy band shift obtained from comparison between experiment and band-structure calculation.

| $x = 0$ | $m^*/m_{band}$ | shift (meV) |
|---|---|---|
| $d_{xy}$ | 10.3 | -27 |
| $d_{yz}$ | 1.4 | -100 |
| $d_{zx}$ | 2.2 | 4 |
| $x = 0.1$ | $m^*/m_{band}$ | shift (meV) |
| $d_{xy}$ | 10.3 | -27 |
| $d_{yz}$ | 1.7 | -82 |
| $d_{zx}$ | 2.1 | 12 |
| $x = 0.2$ | $m^*/m_{band}$ | shift (meV) |
| $d_{xy}$ | 9.8 | -26 |
| $d_{yz}$ | 1.8 | -80 |
| $d_{zx}$ | 2.1 | 12 |
| $x = 0.4$ | $m^*/m_{band}$ | shift (meV) |
| $d_{xy}$ | 9.8 | -16 |
| $d_{yz}$ | 2.3 | -70 |
| $d_{zx}$ | 2.2 | 19 |

Before discussing the mass renormalization in detail, let us discuss the composition dependence of band dispersions. For the Se-substituted compounds, all the three band dispersions around the Γ point are seen more clearly. The $d_{xy}$ (outer band) and $d_{yz}$ (middle band) bands seem to cross the $E_F$ and form two hole Fermi surfaces around the Γ point at the $k_z$ values probed by the measurement using $hv$ = 22 eV photons (as indicated in each panel of Fig. 3). However, the $d_{zx}$ (inner band) remains below $E_F$ at the same $k_z$. Here, the different $k_z$ values for the different compositions are due to the different c-axis lattice parameters [13]. We assumed an inner potential of $V_0$ = 12 eV to calculate $k_z$.) The $d_{zx}$ band is shifted towards the $E_F$ as the Se concentration is increased while the other two bands are not shifted appreciably. From the mass renormalization factors listed in Table 2, one can see that electron correlations are strongly orbital dependent. The $d_{xy}$ band is the most strongly renormalized with the mass renormalization factor of about 10 whereas the other two bands $d_{yz}$ and $d_{zx}$ show moderate mass renormalization factors of 1.5-2. The orbital dependence of mass renormalization is particularly strong for FeTe, where the chalcogen height $h$ is the highest among the iron-based superconductors. The chalcogen height $h$ controls the crystal-field splitting of the Fe 3d orbitals, and the position of the $d_{xy}$ band approaches and crosses the $E_F$ with increasing $h$. The highest $h$ realized in FeTe leads to the strongest crystal-field splitting, the largest $d_{xy}$ character at $E_F$, and hence the strongest electron correlations among the iron-based materials.

In Fig. 4(a), we compare the mass renormalization factors thus obtained with those deduced from the DFT + DMFT calculation [2]. The figure shows that our experimental data agree with theory semi-quantitatively, that is, the $d_{yz}$ and $d_{zx}$ bands show relatively small $m^*/m_{band}$ values, while the $d_{xy}$ band shows a larger $m^*/m_{band}$ value. Our systematic data on the $Fe_{1+y}Te_{1-x}Se_x$ compounds indicate that the Te-rich side of these compounds are particularly strongly correlated materials as compared to the other iron-based superconductors. Our results also confirm the strong orbital dependence of the mass renormalization in $Fe_{1+y}Te_{1-x}Se_x$ when the crystal-field splitting is large. Although Figs. 2 and 3 show the shift of the $d_{zx}$ band with Se concentration, the band shifts listed in Table 2 and plotted in Fig. 4(b) do not show a strong Se concentration dependence because the shift of the $d_{zx}$ band observed in Figs. 2 and 3 mostly originates from the variation of the c-axis parameter and hence of $k_z$ and the strong $k_z$ dispersion of the $d_{zx}$ band.

One of the Fermi surfaces of FeTe consists of the strongly mass-renormalized $d_{xy}$ band, different from the FeSe end member [25], where the $d_{xy}$ band is buried below $E_F$ (at binding energies around 50 meV). This and the unusually large orbital differentiation in FeTe play important roles in the disappearance of superconductivity in the Te-rich region of $FeTe_{1-x}Se_x$ through pair breaking caused by repulsive electron-electron scattering in the $d_{xy}$ band.

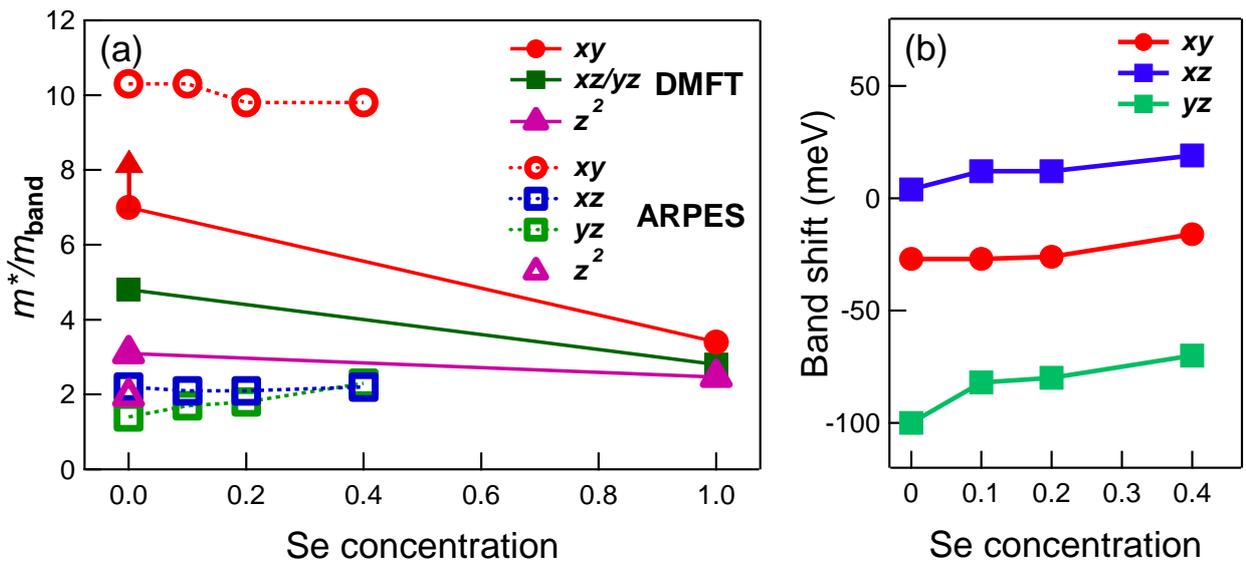

FIG. 4. (a) Comparison of the experimental mass renormalization factors with those from the DFT + DMFT calculations [2]. (b) Energy band shift compared with the DFT calculations as a function of the Se content $x$.#

## Conclusion

We have found that the mass renormalization factors obtained from photoemission measurements for the different bands of $Fe_{1+y}Te_{1-x}Se_x$ (x = 0, 0.1, 0.4, 1) are consistent with the DFT + DMFT calculations [2] and with other experimental data [16, 22]. Our results provide additional evidence for the strong orbital dependence of mass renormalization as well as the strong electron correlation in iron chalcogenides. Our results also provide a systematic set of data for $Fe_{1+y}Te_{1-x}Se_x$ that would help us to resolve and clarify the inconsistencies with previous experimental data [14-19]. Furthermore, the unusually large orbital differentiation of mass renormalization for FeTe and the dominant contribution of the $d_{xy}$ band to the Fermi surface may be the major contributing factors that suppress superconductivity in this compound.

## Acknowledgments
The Stanford Synchrotron Radiation Lightsource is operated by the Office of Basic Energy Science, US Department of Energy. This work was supported by an A3 Foresight Program from the Japan Society for the Promotion of Science. L.C.C.A.II acknowledges support from the Monbukagakusho (MEXT) scholarship through the Embassy of Japan in the Philippines.

#